\documentclass[10pt]{bmc_article}    
\pdfoutput=1

\usepackage{cite} 
\usepackage{url}  
\usepackage{ifthen}  
\usepackage[utf8]{inputenc} 
\usepackage{graphicx}
\usepackage{booktabs}

\newboolean{publ}

\begin{document}

\title{Bayesian inference of
biochemical kinetic parameters using the linear noise approximation}

\author{Micha{\l} Komorowski\correspondingauthor$^{1,2}$%
       \email{Micha{\l} Komorowski\correspondingauthor - m.komorowski@warwick.ac.uk}%
      \and
         B\"arbel Finkenst\"{a}dt$^{1}$%
         \email{B\"arbel Finkenst\"{a}dt\email - B.F.Finkenstadt@warwick.ac.uk}
       \and 
Claire V. Harper$^4$%
         \email{Claire V. Harper - claire.harper@liverpool.ac.uk}%
       \and 
        David A. Rand$^{2,3}$%
 	   \email{David A. Rand - d.a.rand@warwick.ac.uk}%
      }

\address{
\iid(1)Department of Statistics, University of Warwick, Coventry CV4 7AL; \iid(2)Systems Biology Centre, University of Warwick;
\iid(3)Mathematics Institute, University of Warwick; 
\iid(4)Department of Biology, University of Liverpool; }

\maketitle

\begin{abstract}
        \paragraph*{Background:} Fluorescent and luminescent gene reporters 
allow us to dynamically quantify
changes in molecular species concentration
over time on the single cell level. The
mathematical modeling of their interaction
through multivariate dynamical models
requires the development of effective statistical
methods to calibrate such models against
available data. Given the prevalence of stochasticity
and noise
in biochemical systems inference for
stochastic models is of special interest.  In this paper we present a
simple and computationally efficient
algorithm for the estimation of biochemical
kinetic parameters from gene reporter data.      
        \paragraph*{Results:}
We use the linear noise approximation to
model biochemical reactions through a
stochastic dynamic model which essentially
approximates a diffusion model by an ordinary
differential equation model with an
appropriately defined noise process.  An explicit formula
for the likelihood function can be derived
allowing for  computationally  efficient
parameter estimation. The proposed algorithm
is embedded in a Bayesian framework and
inference is performed using Markov chain
Monte Carlo.  
\paragraph*{Conclusions:} The major advantage of the method is that 
in contrast to the more established diffusion
approximation based methods the computationally costly methods of data augmentation are not necessary. Our approach also allows for unobserved variables and measurement error.
The application of the method to both simulated and experimental data
shows that the proposed methodology provides a useful alternative
to diffusion approximation based methods.
\paragraph*{Supplementary Information (SI):} Available at \url{http://www.warwick.ac.uk/staff/M.Komorowski/LNASI.pdf}
\end{abstract}


\section{Background}
The
estimation of parameters in biokinetic models
from experimental data 
is an important problem in Systems Biology.
In general the aim is to calibrate the model
so as to reproduce experimental results in
the best possible way. The solution of this
task plays a key role in interpreting
experimental data in the context of dynamic
mathematical models and hence in
understanding the dynamics and control of
complex intracellular chemical networks and
the construction of synthetic regulatory circuits
\cite{MansEhrenberg11012003}. Among
biochemical kinetic systems, the dynamics
of gene expression and of gene regulatory networks are
of particular interest. Recent developments of
fluorescent microscopy allow us to quantify
changes in protein concentration over time in
single cells (e.g. \cite{MichaelB-Elowitz08162002, Nelson2004})
even with single molecule precision (see
\cite{xie08_single_mol} for review).
Therefore an abundance of data is becoming
available to estimate parameters of
mathematical models in many important cellular systems.

Single cell imaging
techniques have revealed the stochastic
nature of biochemical reactions (see
\cite{JonathanM-Raser09232005,raj2008nno} for review)
that most often occur far from thermodynamic
equilibrium \cite{Keizer1987} and may
involve small copy numbers of reacting
macromolecules \cite{Guptasarma}. This
inherent stochasticity implies that the
dynamic behaviour of one cell is not exactly
reproducible and that there exists stochastic
heterogeneity between cells. The disparate
biological systems, experimental designs and data
types impose conditions on the statistical
methods that should be used for inference
\cite{CarmenG.,Golightly2005,
Fink2008}.
From the modeling point of view the current common 
consensus is that the most exact stochastic description
of the biochemical kinetic system
is provided by the chemical master equation (CME) \cite{Gillespie1992}. 
Unfortunately, for many tasks such as inference the CME is not a convenient
mathematical tool and hence
various types of approximations have been developed. 
The three most commonly used approximations are
\cite{vanKapmen}:
\begin{enumerate}
\item The macroscopic rate equation (MRE)
approach which describes the thermodynamic
limit of the system with ordinary
differential equations and does not take into
account random fluctuations due to the
stochasticity of the reactions.
\item The
diffusion approximation (DA) which provides
stochastic differential equation (SDE) models
where the stochastic perturbation is
introduced by a state dependant Gaussian
noise.
\item The linear noise approximation
(LNA) which can be seen as a combination
because it incorporates the deterministic MRE
as a model  of the macroscopic system and the
SDEs to approximatively describe the
fluctuations around the deterministic
state.
\end{enumerate}

Statistical methods based on the MRE
have been most widely studied
\cite{PMendes11011998,CarmenG.,Ramsay2007,
Esposito}. 
They require
data based on large populations.
The main advantages of
this method are its conceptual simplicity
and the existence of extensive theory for
differential equations.
However, single cells experiments and studies
of noise in small regulatory networks created
the need for statistical tools that are
capable to extract information from
fluctuations in molecular species. Two
methods have been proposed to address this.
The one by \cite{Reinker2006} assumes
availability of single molecule precision
data. Another approach is based on the
diffusion approximation
\cite{Golightly2005, ElizabethA.Heron10012007}.
This uses likelihood
approximation methods (e.g.
\cite{Elerian2001}) that are
computationally intensive and require
sampling from high dimensional posterior
distributions. Inference using these methods is particularly difficult for low frequency data with unobserved model variables \cite{ElizabethA.Heron10012007, Fink2008}. 
The aim of this study is to
investigate the use of the LNA as a method
for inference about kinetic parameters of
stochastic biochemical systems. We find that
the LNA approximation provides an explicit
Gaussian likelihood for models with hidden
variables and measurement error and is
therefore simpler to use and computationally
efficient. To account for prior information
on parameters our methodology is embedded in
the Bayesian paradigm. The paper is
structured as follows: We first provide a
description of the LNA based modeling
approach and then formulate the relevant
statistical framework. We then study its
applicability in four examples, based on both
simulated and experimental data, that clarify
principles of the method. 
\section{Methods}
The chemical master equation (CME) is the
primary tool to model the stochastic
behaviour of a reacting chemical system. It
describes the evolution of the joint
probability distribution of the number of
different molecular species in a spatially
homogeneous, well stirred and thermally
equilibrated chemical system
\cite{Gillespie1992}. Even though these
assumptions are not necessarily satisfied in
living organisms the CME is commonly regarded
as the most realistic model of biochemical
reactions inside living cells. Consider a
general system of $N$ chemical species inside
a volume $\Omega$ and let
$\mathbf{X}=(X_1,\ldots ,X_N)^T$ denote the
number and $\mathbf{x}=\mathbf{X}/\Omega$ the
concentrations of molecules. The
stoichiometry matrix
$\mathbf{S}=\{S_{ij}\}_{i=1,2\ldots N;\
j=1,2\ldots R}$ describes changes in the
population sizes due to $R$ different
chemical events, where each $S_{ij}$
describes the change in the number of
molecules of type $i$ from $X_i$ to $X_i +
S_{ij}$ caused by an event of type $j$. The
probability that an event of type $j$ occurs
in the time interval $[t,t+dt)$ equals
$\tilde{f}_j(\mathbf{x},\Omega,t)\Omega dt$.
The functions
$\tilde{f}_j(\mathbf{x},\Omega,t)$ are called
{\it mesoscopic transition rates}. This
specification leads to a Poisson birth and
death process where the probability
$h(\mathbf{X},t)$ that the system is in the
state $\mathbf{X}$ at time $t$ is described
by the CME \cite{vanKapmen} which is
given in the SI.

The first order terms of a Taylor expansion of the CME
in powers of $1/\sqrt{\Omega}$ are given by
the following MRE (see SI)
\begin{equation}\label{MRE}
\frac{d\phi_i}{dt}=\sum_{j=1}^R
S_{ij}f_j(\varphi,t) \ \ \ \ \ \
i={1,2,\ldots ,N}; \end{equation}
where $
\phi_i=\lim_{\Omega \rightarrow \infty, {X}
\rightarrow \infty}{X}_i/\Omega,\
\varphi=(\phi_1,\ldots ,\phi_N)^T$ and
$f_j(\mathbf{\varphi},t)=\lim_{\Omega
\rightarrow \infty}
\tilde{f}_j(\mathbf{x},\Omega,t).
$\\ 
Including also the second order terms of this
expansion produces the LNA
\begin{equation}\label{DetStochSeparation}
\mathbf{x}(t)=\varphi(t)+\Omega^{-\frac{1}{2}}
\mathbf{\xi}(t)
\end{equation}
which
decomposes the state of the system into a
deterministic part $\mathbf{\varphi}$ as
solution of the MRE in (\ref{MRE}) and a
stochastic process $\mathbf{\xi}$ described
by an It\^{o} diffusion equation
\begin{equation}\label{Ito_LNA}
d\mathbf{\xi}(t)=\mathbf{A}(t)\mathbf{\xi}
dt+\mathbf{E}(t)dW,
\end{equation}
where $dW$
denotes increments of a Wiener process,
$\left[\mathbf{A}(t)\right]_{ik}=\sum_{j=1}^R
S_{ij} {\partial f_j}/{\partial \phi_k}$,
$\left[\mathbf{E}(t)\right]_{ik}=S_{ik}
\sqrt{f_k (\varphi,t)}$ and
$f_i=f_i(\varphi)$ (see SI for derivation).

The rationale behind the expansion in terms of $1/\sqrt{\Omega}$ is that for constant average concentrations  relative fluctuations will decrease with the inverse of the square root of volume \cite{JohanElf11012003}. Therefore the LNA is accurate when fluctuations are sufficiently small in relation to the mean (large $\Omega$).
Hence, the natural measure of adequacy of the LNA
is the coefficient of variation i.e. ratio of the standard deviation to the mean (see SI).
Validity of this approximation is also discussed in details in \cite{JohanElf11012003, Hierarchy}.
In addition it can  be shown that the process describing the deviation from the deterministic state $\Omega^{\frac{1}{2}}(\mathbf{x}-\varphi) $ converges weakly to the diffusion ($\ref{Ito_LNA}$) as $\Omega\rightarrow \infty$ \cite{Kurtz_Realation}.
 In order to use the LNA in a likelihood based inference method we need to derive transition densities of the process $\mathbf{x}$.
\subsection{Transition
densities} The LNA provides solutions that
are numerically or analytically tractable
because the MRE in (\ref{MRE}) can be solved
numerically and the linear SDE in
(\ref{Ito_LNA}) for an initial condition
$\xi(t_{i})=\xi_{t_{i}}$  has a solution of
the form \cite{L.ARNOLD}
\begin{equation}\label{xi_solution}
\xi(t)=\Phi_{t_i}(t-t_i)\left(\xi_{t_i}+\int_{t_i}^t \Phi_{t_i}(s-t_i)^{-1}\mathbf{E}(s)dW(s)\right),
\end{equation}
where the integral is
in the It\^{o} sense and $\Phi_{t_i}(s)$ is
the fundamental matrix of the non-autonomous
system of ODEs
\begin{equation}\label{fundamental}
\frac{d\Phi_{t_i}}{ds}=\mathbf{A}(t_i+s)\Phi_
{t_i},\ \ \ \Phi_{t_i}(0)=I.
\end{equation}
Equations (\ref{xi_solution}),
(\ref{fundamental}) imply that the transition
densities \cite{Oksendal} of the process
$\xi$ are Gaussian\footnote{Throughout the
paper we use  'Gaussian' or 'normal' shortly
to denote either a univariate or a
multivariate normal distribution depending on
the context.}\cite{Oksendal}
\begin{equation}\label{trans_xi}
\mathbf{p}(\xi_{t_i}|\xi_{t_{i-1}},
\Theta)=\psi(\xi_{t_i}|\mu_{i-1},\Xi_{i-1})
\end{equation}
where $\Theta$ denotes a
vector of all model parameters,
$\psi(\cdot|\mu_{i-1},\Xi_{i-1})$ is the
normal density with mean $\mu_{i-1}$ and
variance $\Xi_{i-1}$ specified by
\begin{eqnarray}\label{trans_cov}
\mu_{i-1}&=&\Phi_{t_{i-1}}(\Delta_{i-1})\xi_{
t_{i-1}},\ \ \ \
\Delta_{i-1}=t_i-t_{i-1},\\\nonumber
\Xi_{i-1}&=&\int_{t_{i-1}}^{t_i}(\Phi_{s}(t_i-s)E(s))(\Phi_{s}(t_{i}-s)E(s))^T
ds
\end{eqnarray}
It
follows from (\ref{DetStochSeparation}) and
(\ref{trans_xi}) that the transition
densities of $\mathbf{x}$ are normal
\begin{equation}\label{trans_x}
\mathbf{p}(\mathbf{x}_{t_i}|\mathbf{x}_{t_{i-
1}}, \Theta)=
\psi(\mathbf{x}_{t_i}|\mathbf{\varphi}(t_i)+
\Omega^{-\frac{1}{2}}\mu_{i-1},\Omega^{-1}\Xi_{i-1}).
\end{equation}
The properties of the normal
distribution allow us to derive an explicit
formula for the likelihood of observed 
data.
\subsection{Inference}
It is rarely possible
to observe the time evolution of all
molecular components participating in the
system of interest
\cite{MichalRonen08062002}. Therefore,
we partition the process $\mathbf{x}_t$ into
those components $\mathbf{y}_t$ that are observed
and those $\mathbf{z}_t$ that are unobserved.

Let
 $\bar{\mathbf{x}} \equiv (\mathbf{x}_{t_0},\ \ldots ,
 \mathbf{x}_{t_n})$, $\bar{\mathbf{y}}
\equiv (\mathbf{y}_{t_0},\ \ldots ,\mathbf{y}_{t_n})$
and $\bar{\mathbf{z}} \equiv (\mathbf{z}_{t_0},\
\ldots ,\mathbf{z}_{t_n})$ denote the time-series that
comprise the values\footnote{Here
and throughout the paper we
use the same letter to denote the
stochastic process and its realization.}
of processes $\mathbf{x}$,
$\mathbf{y}$ and $\mathbf{z}$,
respectively, at times $t_0,\ldots t_n$. 

Our aim is to estimate the vector of unknown
parameters $\Theta$ from a sequence of
measurements $\bar{\mathbf{y}}$. 
Given the Markov property of the process
$\mathbf{x}$  the augmented likelihood
$\mathit{P}(\mathbf{\bar{y}},\mathbf{\bar{z
}}|\Theta)$ is given by
\begin{equation}\label{prod_trans_x}
\mathit{P}(\bar{\mathbf{y}},\bar{\mathbf{z}
}|\Theta)=\prod_{i=1}^n \mathbf{p}
(\mathbf{x}_{t_i}|\mathbf{x}_{t_{i-1}},\Theta
)\mathbf{p}(\mathbf{x}_{t_0}|\Theta),
\end{equation}
where
$\mathbf{p}(\mathbf{x}_{t_i}|\mathbf{x}_{t_{i
-1}},\Theta)$ are Gaussian densities
specified in (\ref{trans_x}).
For mathematical convenience we assume that 
$\mathbf{p}(\mathbf{x}_{t_0}|\Theta)$ is also normal with mean $\varphi(t_0)$ and
covariance matrix $\Xi_{-1}$. This assumption is justified as equations (\ref{DetStochSeparation}) and (\ref{Ito_LNA}) imply normal distribution at any time given a fixed initial condition. Mean $\varphi(t_0)$ and covariance matrix $\Xi_{-1}$ are parameterized as elements of $\Theta$.

It can then be
shown that (see SI)
$
\mathbf{\bar{x}} $
is Gaussian. Therefore
\begin{equation}\label{aug_lik}
\mathit{P}(\mathbf{\bar{y}},\mathbf{\bar{z}
}|\Theta)=\psi(\mathbf{\bar{x}}
|
\varphi(t_0),\ldots ,\varphi(t_n),{{\hat{\Sigma
}}}),
\end{equation}
where
$\psi(\cdot|\varphi(t_0),\ldots ,\varphi(t_n),{{\hat{\Sigma
}}})$ is Gaussian
with mean vector
$(\varphi(t_0),\ldots ,\varphi(t_n))$
and covariance matrix ${\hat{\Sigma}}$ whose
elements can be calculated numerically in a
straightforward way (see SI). Since the
marginal distributions are also Gaussian it
follows that the likelihood function
$\mathit{P}(\mathbf{\bar{y}}|\Theta)$ can
be obtained from the augmented likelihood
(\ref{aug_lik})
\begin{equation}\label{normality_XO}
\mathit{P}(\mathbf{\bar{y}}|\Theta)=\psi(\mathbf{\bar{y}}|
(\varphi_y(t_0),\ldots ,\varphi_y(t_n)),{{\Sigma}})
, \end{equation} where the covariance matrix
${\Sigma}=\{{\Sigma}^{(i,j)}
\}_{i,j=0,\ldots ,n}$ is a sub-matrix of
$\hat{\Sigma}$ such that
${\Sigma}^{(i,j)}=Cov(\mathbf{y}_{t_i},\mathbf{y}_{t_j})$
and $\varphi_y$ is the vector consisting of the observed
components of $\varphi$.

Fluorescent reporter data are usually assumed
to be proportional to the number of
fluorescent molecules
\cite{Jian-QiuWu10142005} and
measurements are subject to {\it measurement
error}, i.e.\ errors that do not
influence the stochastic dynamics of the
system. We therefore assume that instead of
the matrix $\mathbf{\bar{y}}$ our data have
the form
$\mathbf{\bar{u}}\equiv\lambda\mathbf{\bar{
y}}+(\epsilon_{t_0},\ldots ,\epsilon_{t_n})$.
The parameter $\lambda$ is a proportionality
constant\footnote{It is straightforward to
generalize for the case with different
proportionality constants for different
molecular components.} and $\epsilon_{t_i}$
denotes a random vector for additive
measurement error. For mathematical
convenience we assume that the joint
distribution of the measurement error is
normal with mean $0$ and known covariance
matrix $\Sigma_\epsilon$, i.e.\
$(\epsilon_{t_0},\ldots ,\epsilon_{t_n}
)\sim N(0,\Sigma_\epsilon)$. If measurement
errors are independent with a constant
variance $\sigma^2_{\epsilon}$ then
$\Sigma_\epsilon=\sigma^2_{\epsilon} I$.
Equation (\ref{normality_XO}) implies that
the likelihood function can be written as
\begin{equation}\label{normality_XM}
\mathit{P}(\bar{{\mathbf{u}}}|\Theta)=\psi(
\bar{{\mathbf{u}}}|
\lambda(\varphi_y({t_0}),\ldots ,\varphi_y (t_n)),\lambda^2{\Sigma
}+\Sigma_{\epsilon}).
\end{equation}
Since
for given data $\bar{{\mathbf{u}}}$ the
likelihood function (\ref{normality_XM}) can
be numerically evaluated any likelihood based
inference is straightforward to implement.
Using Bayes' theorem, the posterior
distribution
$\mathit{P}(\Theta|\bar{\mathbf{u}})$
satisfies the relation \cite{Gamerman}
\begin{equation}\label{posterior}
\mathit{P}(\Theta|\bar{{\mathbf{u}}})\propto
P(\bar{\mathbf{u}}|\Theta)\pi(\Theta).
\end{equation}
We  use the standard
Metropolis-Hastings (MH) algorithm
\cite{Gamerman} to sample from the
posterior distribution in (\ref{posterior}).
 
\section{Results}
In order to study the use
of the LNA method for inference we have
selected four examples which are related to
commonly used quantitative experimental
techniques such as  measurements based on
reporter gene constructs and reporter assays
based on Polymerase Chain Reaction (e.g.
RT-PCR, Q-PCR). For expository reasons, all
case studies consider a model of single gene
expression. %
\subsection{Model of single
gene expression} Although gene expression
involves various biochemical reactions it is
essentially modeled in terms of only three
biochemical species (DNA, mRNA, protein) and
four reaction channels (transcription, mRNA
degradation, translation, protein
degradation)
\cite{MukundThattai07032001,Oude_Clock, ja_BJ}.
Let $\mathbf{x}=(r,p)$ denote concentrations of
mRNA and protein, respectively.
The stoichiometry matrix has the form
\begin{equation}\label{st_matrix_model}
S=\left( \begin{array}{cccc} 1 &-1& 0 &0\\ 0
& 0 &1& -1 \end{array} \right),
\end{equation} where rows correspond to
molecular species and columns to reaction channels. For the reaction
rates
\begin{equation}\label{rates_model}
\mathbf{\tilde{f}}(\mathbf{x})=
(k_R(t),\gamma_R r,k_Pr,
\gamma_P p )^T
\end{equation}
we
can derive the following macroscopic rate
equations \begin{equation}\label{MRE_EX}
\dot{\phi}_R=k_R(t)-\gamma_R\phi_R,\ \ \ \
\dot{\phi}_P=k_P\phi_R-\gamma_P\phi_P.
\end{equation}
For the general case it is assumed
that the transcription
rate $k_R(t)$ is time-dependent, reflecting
changes in the regulatory environment of the gene
such as the availability of transcription factors
or chromatin structure. 

Using (\ref{rates_model}) and 
(\ref{MRE_EX}) in  (\ref{Ito_LNA}) we obtain
the following SDEs describing the deviation
from the macroscopic state
\begin{eqnarray}\label{Ito_xi_SGE}
&&d\xi_R=-\gamma_R\xi_Rdt+\sqrt{k_R(t)+
\gamma_R\phi_R(t)}dW_R,\\\nonumber
&&d{\xi}_P=(k_P\xi_R-\gamma_P{\xi}_P)dt+\sqrt
{k_P\phi_R(t)+\gamma_P{\phi_P}(t)}dW_P.
\end{eqnarray}
We will refer to the model in
(\ref{MRE_EX}) and (\ref{Ito_xi_SGE}) as the
{\it simple model} of single gene expression.

In order to test our method on a nonlinear
system we will also consider the case of an autregulated
network where the transcription rate of the gene is
a function of a modified form of the protein that the gene
codes for and where the modified protein is a 
transcription factor that inhibits the
production of its own mRNA. This is
parameterized by a Hill function
\cite{MukundThattai07032001}
$k_R(t,p)=k_R(t)/(1+({p}/{H})^{n_H})$
where $k_R(t)$ now describes the maximum rate
of transcription, $H$ is a dissociation
constant and $n_H$ is a Hill coefficient.
Thus, the nonlinear autoregulatory model the system is
described by the MRE
\begin{equation}\label{MRE_SGE_AUTO}
\dot{{\phi}}_R=k_R(t,{{\phi}}_P)-\gamma_R{\
phi}_R,\ \ \ \
\dot{{\phi}}_P=k_P\phi_R-\gamma_P{\phi}_P
\end{equation}
and the SDEs
\hspace{-5mm}
\begin{eqnarray}\nonumber
d\xi_R&=&(k'_R(t)
\xi_P\ -\ \gamma_R\xi_R)dt\ \ \ +\ \ \ \sqrt{k_R(t)+\gamma_R
\phi_R(t)}dW_R\\\label{Ito_xi_SGE_AUTO}
d{\xi}_P&=&(k_P\xi_R-\gamma_P{\xi}_P)dt+\sqrt{
k_P\phi_R(t)+\gamma_P{\phi_P}(t)}dW_P
\end{eqnarray}
where $k'_R(t)\equiv
{\partial k_R(t,{{\phi}}_P )}/{\partial
\phi_P}$. 
We refer to this
model as {\it the autoregulatory model} of single
gene expression. The two models constitute
the basis of our inference studies below.

\subsection{Inference from fluorescent
reporter gene data for the simple model of
single gene expression}
To test the algorithm
we first use the simple model of single gene
expression. We generate data according to the
stoichiometry matrix (\ref{st_matrix_model})
and rates (\ref{rates_model}) using
Gillespie's algorithm
\cite{Gillespie1977} and sample it at
discrete time points. We then generate
artificial data that are proportional to the
simulated protein data with added normally
distributed measurement error with known
variance $\sigma_{\epsilon}^2$. Furthermore
we assume that mRNA levels are unobserved.
Thus the data are of the form\footnote{The
volume of the system $\Omega$ is unknown and
we set $\Omega=1$ so that concentration
equals the number of molecules.}
\begin{equation}\label{pM}
\bar{\mathbf{u}}=({{u}}_{t_0},\ldots ,{u}_{
t_n})^T , \end{equation}
where
$u_{t_i}=\lambda p_{t_i}+\epsilon_{t_i}$,
$p_{t_i}$ is the simulated protein concentration,
$\lambda$ is an unknown proportionality
constant and $\epsilon_{t_i}$ is measurement
error. %
For the purpose of our
example we model the transcription
function
by
\begin{equation}\label{time_dep_trans}
k_R(t)=\Bigg\{ \begin {array}{ccccc} b_0
\exp(-b_1(t-b_3)^2) + b_4 \ \ \ \ \ \ \ t\leq
b_3\\\noalign{\medskip} b_0
\exp(-b_2(t-b_3)^2) + b_4 \ \ \ \ \ \ \ t >
b_3\end{array}
\end{equation}
This form of
transcription corresponds to an experiment,
where transcription increases for $t \leq
b_3$ as a result of being induced by an
environmental stimulus and for $t > b_3$
decreases towards a baseline level $b_4$.\\ We
assume that at time $t_0$ ($t_0<<b_3$) the
system is in a stationary state. Therefore,
the initial condition of the MRE is a
function of unknown parameters
$(\phi_R(t_0),{\phi}_P(t_0))=({b_4}/{
\gamma_R},{b_4k_P}/{\gamma_R\gamma_P}).$

To ensure identifiability of all model
parameters we assume that informative prior
distributions for both degradation rates are
available. Priors for all other parameters
were specified to be non-informative. 

To
infer the vector of unknown parameters
$$
\Theta=(\gamma_R, \gamma_P, k_P, \lambda,
b_0, b_1, b_2, b_3, b_4)
$$
we sample from the
posterior distribution
$$\mathit{P}(\Theta|\bar{\mathbf{u}})
\propto
\mathit{P}(\bar{\mathbf{u}}|\Theta)\pi(
\Theta)$$
using the standard MH algorithm. The
distribution
$\mathit{P}(\bar{\mathbf{u}}|\Theta)$ is
given by (\ref{normality_XM}). 

The protein level of the simulated trajectory
is sampled every $15$ minutes
and a sample size of $101$ points obtained.
We
perform inference for two simulated data
sets: estimate 1 is based on a single
trajectory while estimate 2 represents a
larger data set using 20 sampled trajectories
(see Figure 1A). All
prior specifications, parameters used for the
simulations and inference results are
presented in Table 1A.

Estimate 1 demonstrates that it is possible to infer all
parameters from a single, short length time series
with a realistically achievable time resolution.
Estimate 2 shows that usage of the LNA
does not seem to result in any significant bias.
A bias has not been detected despite the very
small number of mRNA molecules
(5 to 35 - Figure 2A in the SI) and
protein molecules (100 to 500 - Figure
1A). 
The coefficient of variation varied between
approximately 0.15 and 0.4 for both
molecular species  (Figure 1 in the SI).

Inference for this model required sampling
from the 9 dimensional posterior distribution
(number of unknown parameters).
If instead one used a
diffusion approximation based
method it would be necessary to sample from
a posterior distribution of much
higher dimension (see SI). In addition,
incorporation of the measurement error is
straightforward here, whereas for other
methods it involves a substantial computational
cost \cite{ElizabethA.Heron10012007}.
\begin{table}
{\begin{tabular}{lllll}
\toprule
{\bf (A)} & & & & \\
Param.&Prior&Value&Estimate 1&Estimate 2\\\midrule $\gamma_R$&$\Gamma$(0.44,$10^{-2}$)&~0.44&0.43 (0.27-0.60)&~0.49 (0.38-0.61)\\
$\gamma_P$&$\Gamma$(0.52,$10^{-2}$)&~0.52&0.51 (0.35-0.67)&~0.49 (0.38-0.61)\\ $k_P$&Exp(100)&10.00&21.09 (3.91-67.17)&~11.41 (7.64-16.00)\\
$\lambda$&Exp(100)&~1.00&1.42 (0.60-2.57)&~1.08 (0.76-1.36)\\ $b_0$&Exp(100)&15.00&6.80 (0.97-18.43)&12.78 (9.80-15.33)\\ $b_1$&Exp(1)&~0.40&~0.79 (0.05-3.02)&~0.29 (0.18-0.43)\\
$b_2$&Exp(1)&~0.40&0.77 (0.08-2.79)&~0.77 (0.32-1.59)\\
$b_3$&Exp(10)&~7.00&6.13 (4.41-7.85)&~7.25 (6.79-7.55)\\
$b_4$&Exp(100)&~3.00&0.94 (0.11-2.88)&~2.87 (2.11-3.52)\\ \midrule
{\bf (B)} & & & & \\
Param.&Prior&Value&Estimate 1&Estimate 2\\\midrule $\gamma_R$&$\Gamma$(0.44,$10^{-2})$&~~0.44&~~0.44 (0.27-0.60)&0.42 (0.32-0.54)\\
$\gamma_P$&$\Gamma$(0.52,$10^{-2})$&~~0.52&~~0.49 (0.33-0.65)&0.49 (0.36-0.61)\\ $k_P$&Exp(100)&~10.00&~~14.86 (3.18-47.97)&9.35 (5.87-13.15)\\
$\lambda$&Exp(100)&~~1.00&~~1.26 (0.48-2.30)&1.15 (0.81-1.50)\\ $b_0$&Exp(100)&~15.00&~5.99 (0.20-23.06)&13.47 (9.24-17.13)\\
$b_1$&Exp(1)&~~0.40&~~0.59 (0.01-2.75)&0.27 (0.14-0.53)\\ $b_2$&Exp(1)&~~0.40&~~0.92 (0.05-2.92)&0.83 (0.21-3.52)\\ $b_3$&Exp(10)&~~7.00&~~6.53(0.74-14.69)&7.27 (6.44-7.79)\\ $b_4$&Exp(100)&~~3.00&~~2.85 (0.35-7.19)&2.64 (1.82-3.32)\\
\midrule
\end{tabular}}
\caption{Inference results for {\bf (A)} the simple  model  and {\bf (B)} autoregulatory model of single gene expression\label{TABLE1} 
Parameter values used in simulation, prior distribution, posterior medians and 95\% credibility intervals.
Estimate 1 corresponds to inference from single time series, Estimate 2 relates to estimates obtained from 20 independent time series. Data used for
inference are plotted in Figure \ref{PIC_DATA_GFP_AGFP}A for case {\bf A} and Figure \ref{PIC_DATA_GFP_AGFP}B for case {\bf B}. Variance of the measurement error was assumed to be known $\sigma_{\epsilon}=9$. Rates
are per hour. Parameters are $n_H=1$, $H=61.98$ in case {\bf B}.
}
\end{table} 

\begin{figure}[!t]
\begin{center}
  \includegraphics[scale=0.28]{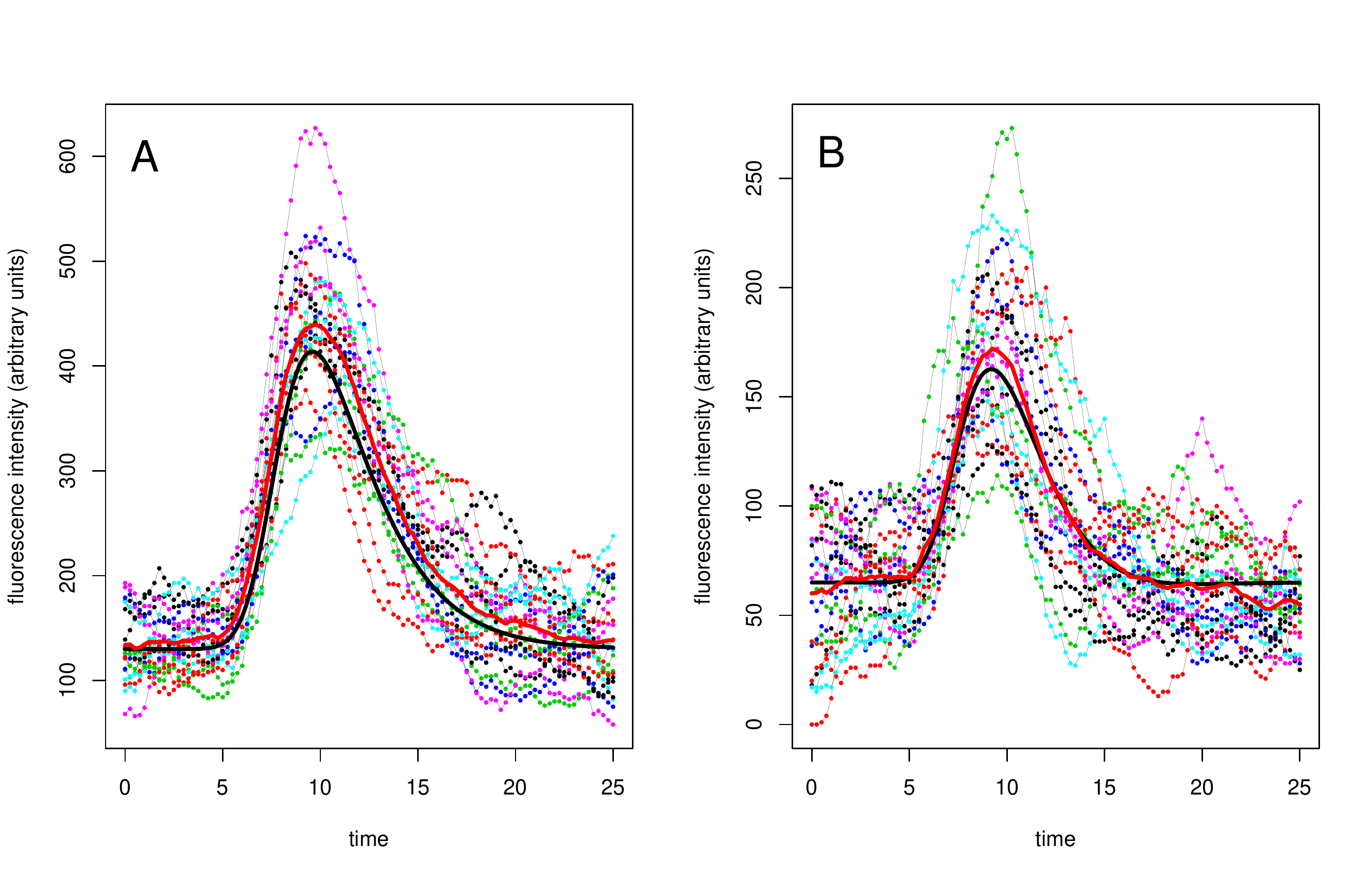}
\end{center}
   \caption{
Protein timeseries generated using Gillespie's
algorithm for the simple {\bf{A}} and autoregulatory
{\bf{B}} models of single gene expression with added
measurement error $(\sigma^2_{\epsilon}=9)$.
Initial conditions for mRNA and protein were
sampled from Poisson distributions with means equal
to the stationary means of the system with equal
constant transcription rate $b_4$. In the
autoregulatory case we set
$H={b_4k_P}/{2\gamma_R\gamma_P}$. In each panel
20 time series are presented. The deterministic and
average trajectories are plotted in bold solid and dashed lines respectively.
 Corresponding mRNA trajectories (not
used for inference) are presented in the SI.
\label{PIC_DATA_GFP_AGFP}}
\end{figure}

\subsection{Inference from fluorescent reporter gene data for the model of single gene expression with autoregulation}
The following example considers the autoregulatory
system with only a small number of reacting
molecules. Using Gillespie's algorithm we generate
artificial data from the single gene expression
model with autoregulation. The protein time courses
were then sampled every 15 minutes at 101 discrete
points per trajectory (see Figure
1B). As before we assume that
the mRNA time courses are not observed and that the
protein data are of the form given in
(\ref{pM}), i.e.\ proportional to the actual amount
of protein  with additive Gaussian measurement
error. As in the previous case study we estimate
parameters from two simulated data sets,  a single
trajectory and an ensemble of 20 independent
trajectories. The inference results summarized in
Table 1B show that despite the low
number of mRNA (0-15 molecules, see  Fig. 2 in SI)
and  protein (10-250 molecules, see Fig.
\ref{PIC_DATA_GFP_AGFP}B) all parameters can be
estimated well with appropriate precision.

\subsection{Inference for PCR based reporter data}
In the case of reporter assays based on Polymerase
Chain Reaction (e.g. RT-PCR, Q-PCR) measurements
are obtained from the extraction of the molecular
contents from the inside of cells. Since the sample
is sacrificed, the sequence of measurements are not
strictly associated with a stochastic process
describing the same evolving unit. Assume that at
each time point $t_i$ $(i=0,..n)$ we observe $l$
measurements that are proportional to
the number of RNA molecules either from a single cell
or from a population of $s$ cells. This gives a
$(n+1)\times l$ matrix of data points
\begin{equation}\label{QPCR_data}
\mathbf{\bar{u}}\equiv\left\{{u}_{t_i,j}\right\}_{i=0,\ldots
n;j=1,\ldots ,l} \end{equation}
where
${u}_{t_i,j}=\lambda
{r}_{t_i,j}+\epsilon_{t_i,j}$, ${r}_{t_i,j}$ is the
actual RNA level, $\lambda$ is the
proportionality constant, $\epsilon_{t_i,j}$ is a
Gaussian independent measurement error indexed by
time $t_i$; $j=1,\ldots ,l$ indexes the $l$
measurements that are taken at
time $t_i$. Note that
 ${r}_{t_i,j}$ and ${r}_{t_{i+1},j}$ are
independent random variables as they refer to
different cells. We assume that the dynamics of RNA
is described by the simple model of single gene
expression with LNA equations (\ref{MRE_EX}) and
(\ref{Ito_xi_SGE}).  Let $\Upsilon_{t}$
denote the distribution
of measured RNA at time $t$ $({u}_t\sim\Upsilon_{t})$. In order to
accommodate for the different form of data we 
modify the estimation procedure as follows. For
analytical convenience we assumed that the initial
distribution is normal
$\Upsilon_{t_0}=N(\mu_{t_0},\sigma^2_{t_0})$. This
together with eq. (\ref{trans_x}) and normality of
measurement error implies that
$\Upsilon_{t}=N(\mu_t,\sigma^2_t)$. Simple explicit
formulae for $\mu_t$ and $\sigma^2_t$ are derived
in the SI. Since all observations
${u}_{t_i,\cdot}$ are independent we can write
the posterior distribution as
\begin{equation}
\pi(\Theta|\mathbf{\bar{u}})\propto \prod_{i=0}^{n}
\prod_{j=1}^l
\psi(u_{t_i,j}|\mu_{t_i},\sigma^2_{t_i})\
\pi(\Theta), \end{equation}
where
$\psi(\cdot|\mu_{t_i},\sigma^2_{t_i})$ is the
normal density with parameters
$\mu_{t_i},\sigma^2_{t_i}$. In order to infer the
vector of the unknown parameters $\Theta=(\gamma_R,
\lambda, b_0, b_1, b_2, b_3,
b_4,\mu_{t_0},\sigma^2_{t_0})$ we sample from the
posterior using a standard MH algorithm. To test
the algorithm we have simulated a small ($l=10$,
$n=50$, plotted in Figure 2) and
a large ($l=100$, $n=50$) data set using
Gillespie's algorithm with parameter values given
in Table 2. The data were sampled
discretely every $30$ minutes and  a standard
normal error  was added. Initial conditions were
sampled from the Poisson distribution with mean
$b_4/\gamma_R$. The estimation results in Table
2 show that parameters can be inferred
well in both cases even though the number of RNA
molecules in the generated data is very small
(about 5-35 molecules). Since subsequent
measurements do not belong to the same stochastic
trajectory, estimation for the model presented here
is not straightforward  with the diffusion
approximation based methods.

\begin{figure}[!t]
\begin{center}\includegraphics[width=0.23\textwidth,totalheight=0.2\textheight]{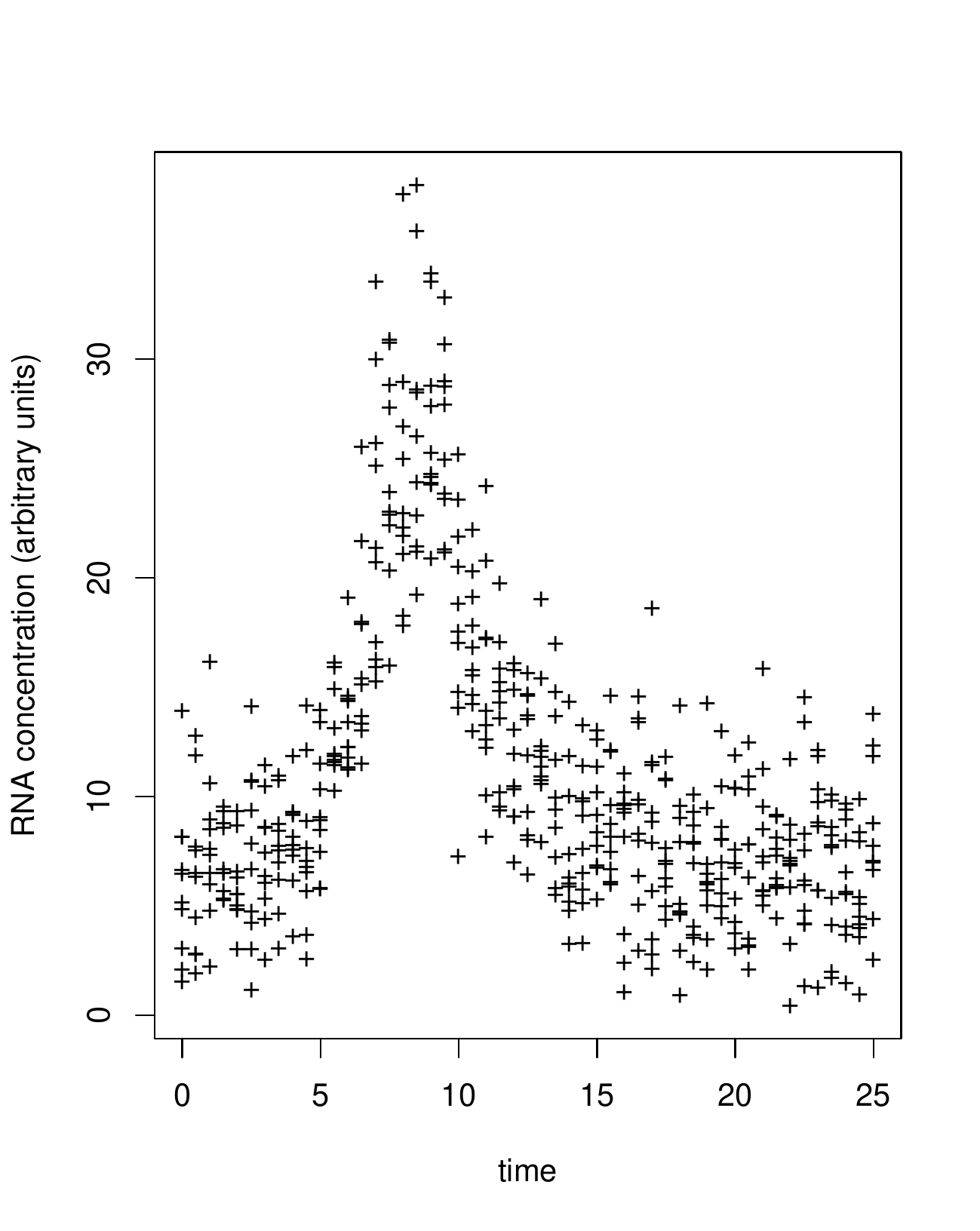}
\includegraphics[width=0.23\textwidth,totalheight=0.2\textheight]{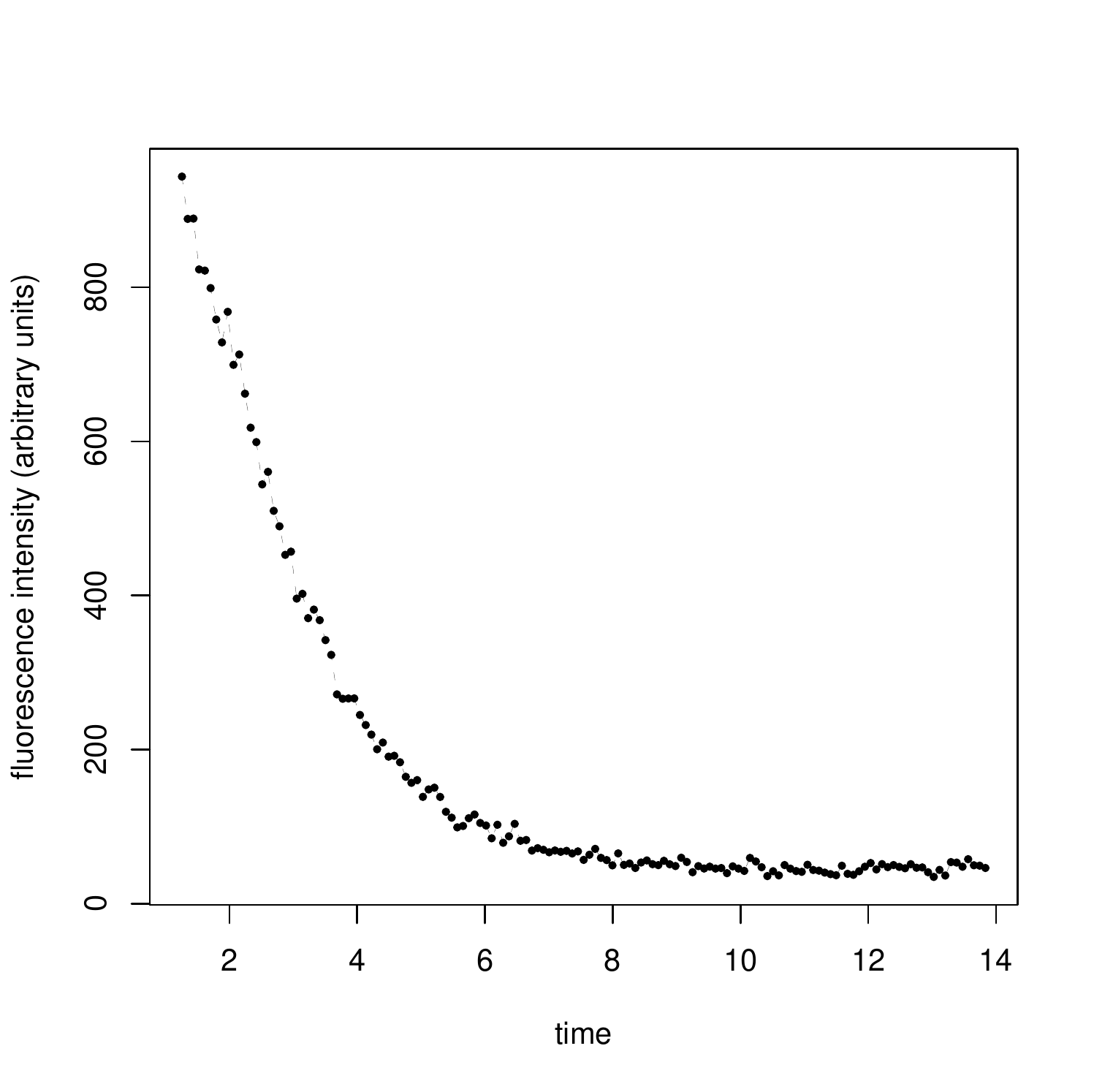}\end{center}
   \caption{
{\bf{Left:}}PCR based reporter assay data simulated
with Gillespie's algorithm using parameters
presented in Table \ref{TABLE2} and extracted $51$
times (n=50), every $30$ minutes {with 
an independently and normally distributed error
($\sigma^2_{\epsilon}=9)$}. Each cross correspond
to the end of simulated trajectory, so that the
data drawn are of form (\ref{QPCR_data}). Since
number of RNA molecules is know upto
proportionality constant y-axis is in arbitrary
units. Time on x-axis is expressed in hours.
Estimates inferred form this data are shown in
column {\it Estimate 1} in Table \ref{TABLE2}.
{\bf{Right:}} Fluorescence level from cycloheximide
experiment is plotted against time (in hours).
Subsequent dots correspond to measurements taken
every 6 minutes. \label{PIC_DATA_QPCR} }
\end{figure}

\begin{table}
\begin{tabular}{lllll}\toprule
Parameter&Prior&Value&Estimate 1&Estimate 2\\\midrule 
$\gamma_R$&Exp(1)&~0.44&~0.45 (0.35-0.60)&~0.46 (0.42-0.50)\\ 
$\lambda$&Exp(100)&~1.00&~1.07 (0.90-1.22)&~1.01 (0.95-1.05)\\ 
$b_0$&Exp(100)&15.00&13.13 (10.20-15.87)&14.91 (13.86-15.77)\\ 
$b_1$&Exp(1)&~0.40&~0.29 (0.14-0.51)&~0.43 (0.32-0.54)\\ 
$b_2$&Exp(1)&~0.40&~0.32 (0.12-0.93)&~0.32 (0.21-0.43)\\ 
$b_3$&Exp(10)&~7.00&~7.05 (6.39-7.63)&~6.99 (6.76-7.15)\\ 
$b_4$&Exp(100)&~3.00&~2.97 (2.00-4.18)&~3.10 (2.76-3.43)\\ 
$\mu_0 $&Exp(100)&~6.76&~6.90 (5.79-7.69)&~6.55 (6.14-6.85)\\ 
$\sigma^2_0$&Exp(100)&~6.76&~3.52 (0.01-8.99)&~7.59 (5.44-9.49)\\ 
\midrule
\end{tabular}
\caption{Inference results for PCR based reporter assay simulated data\label{TABLE2}
Parameter values used to generate data, prior
distributions used for estimation,  posterior
median estimates together with 95\% credibility
intervals. Estimate 1, Estimate 2 columns relate to
small (l=5, n=50) and large (l=100, n=50) sample
sizes. Variance of the measurement was assumed to
be known $\sigma^2_{\epsilon}=4$. Estimated rates
are per hour.}
\end{table}

\subsection{Estimation of gfp protein degradation
rate from cycloheximide experiment}
In this section the method is applied to
experimental data. After a period of
transcriptional induction, translation of gfp was
blocked by the addition of cycloheximide (CHX). Details of the experiment are presented in the SI.
Fluorescence was imaged every 6 minutes for 12.5h
(see Figure 2). Since inhibition
may not be fully efficient we assume that
translation may be occurring at a (possibly small)
positive rate $k_P$. The model with the LNA is
\begin{eqnarray}
\dot{{\phi}}_P&=&k_P-\gamma_P{\phi}_P,\\\nonumber
d{\xi}_P&=&-\gamma_P\xi_Pdt+\sqrt{k_P+\gamma_P{\phi}_P}dW_{P}.
\end{eqnarray}
The observed fluorescence is assumed
to be proportional to the signal with
proportionality constant $\lambda$. For comparison
we also consider  the diffusion approximation for
which an exact transition density can be derived
analytically (see SI for derivation)
\begin{eqnarray} dp=
(k_P-\gamma_Pp)dt+\sqrt{k_P+\gamma_P{p}}dW_{P}.
\end{eqnarray}
Since incorporation of  measurement
error for the diffusion approximation based model
is not straightforward, we assume that measurements were taken
without any error to ensure fair comparison between the two approaches. Table
3 shows that estimates obtained with
both methods are very similar.
\begin{table}
{\begin{tabular}{llll}\toprule Param.&Prior&Estimate LNA&Estimate
DA\\\midrule $\gamma_P$&Exp(1)& 0.45 (0.31-0.62)& 0.53 (0.39-0.67)\\ $k_P$&Exp(50)&0.32(0.10-1.75)&0.43 (0.16-1.07)\\ $\lambda$&Exp(50)&
22.79(13.79-36.92)& 23.85(16.31-36.54)\\ $\tilde{\phi}_P(0)$&$N(u_{t_0},u_{t_0})$& 889.03(831.44-945.34)& -\\\midrule
\end{tabular}}
\caption{Inference results for CHX experimental data\label{TAB_DEG}.
Priors, posterior mean and 95\% credibility
intervals obtained from CHX experimental data using
the LNA approach and diffusion approximation
approach. Estimation with the LNA involved one more
parameter $\tilde{\phi}_P(0)=\lambda {\phi}_P(0)$.
Estimated rates are per hour.}
\end{table}

\section{Discussion}
The aim of this paper is to suggest the LNA as a
useful and novel approach to the inference of
biochemical kinetics parameters. Its major
advantage is that an explicit formula for the
likelihood can be derived even for systems with
unobserved variables and data with additional
measurement error. In contrast to the more established
diffusion approximation based methods
\cite{Golightly2005,ElizabethA.Heron10012007}
the computationally
costly methods of data augmentation to approximate
transition densities and to integrate out
unobserved model variables are not necessary.
Furthermore, this method can also accommodate measurement
error in a straightforward way. The suggested
procedure here is implemented in a Bayesian
framework using MCMC simulation to generate
posterior distributions. The  LNA has previously
been studied  in the context of approximating
Poisson birth and death processes
\cite{Kurtz_Realation,JohanElf11012003,Tomioka,
Hierarchy} and it was shown that for a
large class of models the LNA provides an excellent
approximation. Furthermore, in \cite{Tomioka}
it is shown that for the systems with linear reaction
rates the first two moments of the transition
densities resulting from the CME and the LNA are
equal. Here we propose using the LNA directly for
inference and provide evidence that the resulting
method can give very good results even if the
number of reacting molecules is very small. Our
experience from previous works with diffusion
approximation based methods  \cite{Fink2008,
ElizabethA.Heron10012007} is that their
implementation is challenging especially for 
data that are sparsely sampled in time because the
need for imputation of unobserved time points leads
to a very high dimensionality of the posterior
distribution. This usually results in highly
autocorrelated traces affecting the speed of
convergence of the Markov chain. Our method
considerably reduces the dimension of the posterior
distribution to the number of unknown parameters of
a model only and is independent of the number of
unobserved components. Nevertheless it can only be
applied to the systems with sufficiently large
volume, where  fluctuations around a deterministic
state are relatively close to the mean.

\section{Authors contributions}
MK proposed and implemented the algorithm. 
CVH performed the cycloheximide experiment. MK wrote the paper with assistance from BF and DAR, who supervised the study.     

{\ifthenelse{\boolean{publ}}{\footnotesize}{\small}

 \bibliographystyle{bmc_article}  
  \bibliography{tau}     

}

\end{document}